\documentclass[conference]{IEEEtran}
\makeatletter
\def\ps@headings{%
\def\@oddhead{\mbox{}\scriptsize\rightmark \hfil \thepage}%
\def\@evenhead{\scriptsize\thepage \hfil \leftmark\mbox{}}%
\def\@oddfoot{}%
\def\@evenfoot{}}
\makeatother
\usepackage[linesnumbered,boxed]{algorithm2e}

\usepackage{epsfig}
\usepackage{fixltx2e}
\usepackage{flushend}
\usepackage{url}
\usepackage{amsmath}
\usepackage[utf8]{inputenc}
\usepackage{amssymb}
\usepackage{graphicx}
\usepackage{subfig}

\newcommand{\BfPara}[1]{{\noindent {\bf #1}}}
\newcommand{\ignore}[1]{}
\DeclareMathOperator*{\argmin}{\arg\!\min}

\begin{document}

\title{Honey Onions: a Framework for Characterizing and Identifying
  Misbehaving Tor HSDirs}

\author{\IEEEauthorblockN{Amirali Sanatinia, Guevara Noubir}
\IEEEauthorblockA{College of Computer and Information Science\\
Northeastern University, Boston, USA\\
\{amirali,noubir\}@ccs.neu.edu}
}

\maketitle

\begin{abstract}

In the last decade, Tor proved to be a very successful and widely popular system to protect users' anonymity. However, Tor remains a practical system with a variety of limitations, some of which were indeed exploited in the recent past. In particular, Tor's security relies on the fact that a substantial number of its nodes do not misbehave. In this work we introduce, the concept of honey onions, a framework to detect misbehaving Tor relays with HSDir capability. This allows to obtain lower bounds on misbehavior among relays. We propose algorithms to both estimate the number of snooping HSDirs and identify the most likely snoopers. Our experimental results indicate that during the period of the study (72 days) at least 110 such nodes were snooping information about hidden services they host. We reveal that more than half of them were hosted on cloud infrastructure and delayed the use of the learned information to prevent easy traceback.

\end{abstract}

\section{Introduction}
\label{s:intro}

Over the last decade, Tor emerged as a popular tool and infrastructure
that protects users' anonymity and defends against tracking and
censorship. It is used today by millions of ordinary users to protect
their privacy against corporations and governmental agencies, but also
by activists, journalists, businesses, law enforcements and
military~\cite{tor-stats}. 

The success and popularity of Tor makes it a prime target for
adversaries as indicated by recent
revelations~\cite{snowden-tor-stinks}. Despite its careful design,
that significantly improved users privacy against typical adversaries,
Tor remains a practical system with a variety of limitations and
design vulnerabilities, some of which were indeed exploited in the
past~\cite{CMU-FBI-Tor,trawltor}. Due to the perceived security that
Tor provides, its popularity, and potential implication on its users,
it is important that the research community continues analyzing and
strengthening its security.



\begin{figure}
\centering
\includegraphics[width=0.45\textwidth]{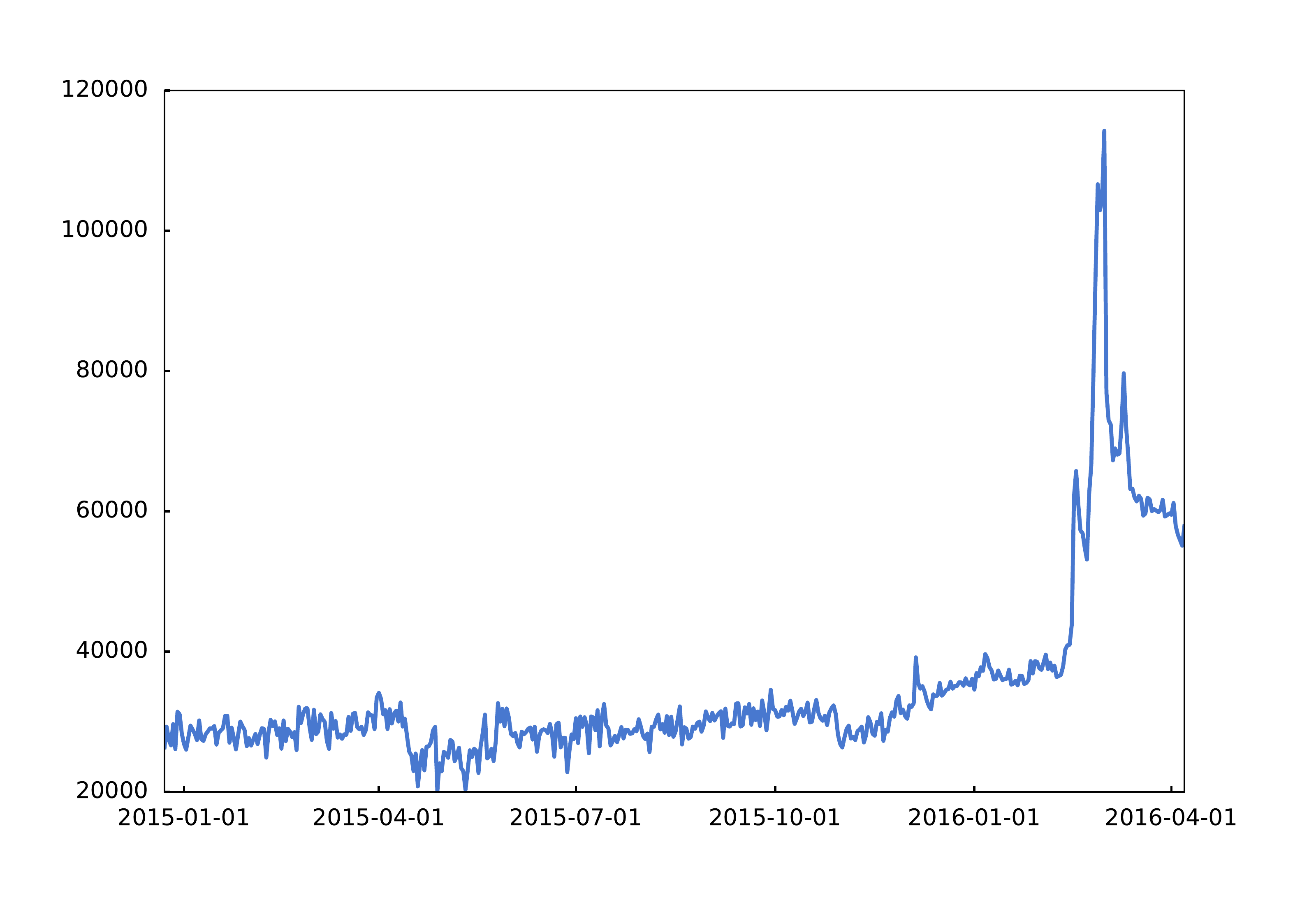}
\caption{Recent unexplained surge in the number of Hidden
  Services. The number of hidden services (\texttt{.onion}) suddenly
  tripled, before settling at twice the number before the surge.}
\label{f:hs_stat}
\end{figure}

This is specially important since users typically have a poor
understanding of the privacy protection that Tor really provides as
evidenced by past events. For instance, in a highly publicized case,
security researchers collected thousands of sensitive e-mails and
passwords from the embassies of countries including India and
Russia~\cite{Tor-embassies}. These embassies used Tor believing it
provides end-to-end encryption, sending sensitive un-encrypted data
through malicious exit nodes. Other research revealed that many users
run BitTorrent over Tor, which is insecure and resulted in
deanonymization~\cite{Tor-BitTorrent}. Finally, recent incidents
revealed that the Tor network is continuously being attacked by a
variety of organizations from universities to governmental agencies,
with difficult to predict
ramifications~\cite{CMU-FBI-Tor,Deanonymization}. Even more recently,
the still unexplained sudden surge in the number of hidden services
(\texttt{.onion}), more than tripling their number before returning to
relatively smaller numbers (See Figure~\ref{f:hs_stat}), indicates
that the Tor network is not well understood, in part due to its
peer-to-peer nature, the privacy services it provides that limit
measurements, and the attacks that it attracts~\cite{onionbots}.

Tor's security, by design, relies on the fact that a substantial
number of its relays should not be malicious. It is however difficult
to assess to what extent this condition holds true. The fact that many
attacks are passive, makes it even harder to assess the significance
of this threat. In this work, we developed a framework, techniques,
and a system to provide some elements of the answer to this
challenging problem.
  
We introduce the concept of \textit{honey onions} (\texttt{honions}),
to expose when a Tor relay with Hidden Service Directory (HSDir) capability has been modified to
snoop into the hidden services that it currently hosts. We believe
that such a behavior is a clear indicator of sophisticated malicious
activity, for it not only is explicitly undesired by the Tor
Project~\cite{tor-project-talk} but also requires a modification to
the Tor software, indicating some level of sophistication of the
perpetrator. Honions are hidden services that are created for the sole
purpose of detecting snooping, and are not shared or publicized in any
other form. Therefore, any visits on the server side of the honion is
a clear indication that one of the HSDir that hosted it is
snooping. Since hidden services are hosted on multiple HSDirs and
change location on a daily basis, it is not easy to infer which HSDir
is the malicious\footnote{In this paper, we use the terms malicious
  and snooping interchangeably.} one. The visits information leads to
a bipartite graph connecting honions and HSDirs. Finding the smallest
subset of HSDirs that can explain honion visits provides a lower-bound
on the number of malicious HSDirs. This has the benefit of giving a
sense of the scale of malicious behavior among Tor relays. We show
that this problem can be formulated as a Set Cover, an NP-Complete
problem. We develop an approximation algorithm to this specific
problem as well as an Integer Linear Program (ILP) formulation. We
build a system to deploy the honions along with a schedule for the
lifetime of each one of them to maximize the collected information
without generating an excessive number of hidden services. The
generated honions have a lifetime of one day, one week, or one
month. Throughout the experiment, which lasted 72 days so far, the
maximum number of generated honions did not exceed 4500 hidden
services (which is significantly lower than the anomaly that hidden
services are experiencing). Based on the experimental data, we are
able to infer that there are at least 110 snooping HSDirs. A careful
analysis of the experimental data and results from the ILP solution,
allows us to infer most of the misbehaving HSDirs and their most likely
geographical origin. Based on these results we are able to classify
misbehaving HSDirs in two main categories, \textit{immediate
  snoopers}, and \textit{delayed snoopers}.  \ignore{
  \textit{Persistent-Immediate Snoopers}, \textit{Persistent-Delayed
    Snoopers}, \textit{Randomized Deterministic-Delay Snoopers}, and
  \textit{Randomized Deterministic-Delay Snoopers}.} Immediately and
deterministically visiting a honion results in a higher detection and
identification. However, delaying and randomization reduces the
traceability (as other HSDir who hosted the honion could also be
blamed) at the expenses of potentially missing key information that
the hidden service creator might put for only a short period of
time. Therefore, a smart HSDir snooper has to trade-off delay (and
risk of missing information) with risk of detection. In this paper, we
discuss the behavior and characteristics of the malicious HSDirs. We
found out that more than half the malicious HSDirs are of the delayed
type, and are hosted on cloud infrastructure. Our contributions can be
summarized as follows:

\begin{itemize}

\item The honey onion framework for detecting snooping HSDirs.

\item An approximation algorithm and Integer Linear Program for
  estimating and identifying the most likely snooping HSDirs.

\item An experimental study leading to the discovery of at least 110
  snooping HSDirs and a peek into their behavior.

\end{itemize}

\ignore{This naturally leads to a game-theoretic
formulation of the interactions between the honions system and
snooping HSDirs.}

The rest of the paper is structured as follows. In Section~\ref{s:tor}
we overview the architecture of Tor hidden services and
HSDirs. Section~\ref{s:approach} outlines our approach, and system
architecture. Section~\ref{s:detection} provides the formalization of
the detection and identification problem, shows the reduction to the
set cover problem, and the approximation algorithm as well as the
Integer Linear Programming formulation. In Section~\ref{s:experiment}, we
discuss our implementation of the system, report on the experimental
results when processed by the identification algorithms. In
Section~\ref{s:discussion} we discuss the experimental results and the
characteristics and behavior of malicious HSDirs. In
Section~\ref{s:related} we summarize the prior and related work. We
conclude the paper in Section~\ref{s:conclusion}.
\section{Hidden Services \& Hidden Service Directories}\label{s:tor}

Tor~\cite{Tor} is an anonymity network that allows users to circumvent
censorship and protect their privacy, activities and location from
government agencies and corporations. Tor also provides anonymity for
the service provided with hidden services, which enables them to
protect their location (IP address), yet allowing users to
connect to them. Hidden services have been used to protect both
legitimate and legal services for privacy conscious users (e.g.,
Facebook), and for illicit purposes such as drug and
contraband market~\cite{Christin:2013:silkroad, silk2}, and
extortion. This attracts attacks from a variety of
actors. In order to understand the specific HSDirs snooping
misbehavior we are interested and the honion system setup and
algorithms, we first summarize some key mechanisms of Tor.  In
particular, we focus on the architecture of hidden services, both from
the client and the service provider perspective.

The Tor hidden services architecture is composed of the following
components:

\begin{itemize}
\item{\emph{Server}, that runs a service (e.g., a web server).}
\item{\emph{Client}, that wishes to access the server.}
\item{\emph{Introduction Points (IP)}, a set of Tor relays, chosen by
  the hidden service, that forward the initial messages between the
  server and the client's Rendezvous Point.}
\item{\emph{Rendezvous Point (RP)}, a Tor relay randomly chosen by the
  client that forwards the data between the client and the hidden
  service.}
\item{\emph{Hidden Service Directories (HSDir)}, a set of Tor relays
  chosen by the server to store its descriptors.}
\end{itemize}

\begin{figure}
\centering
\includegraphics[width=0.45\textwidth]{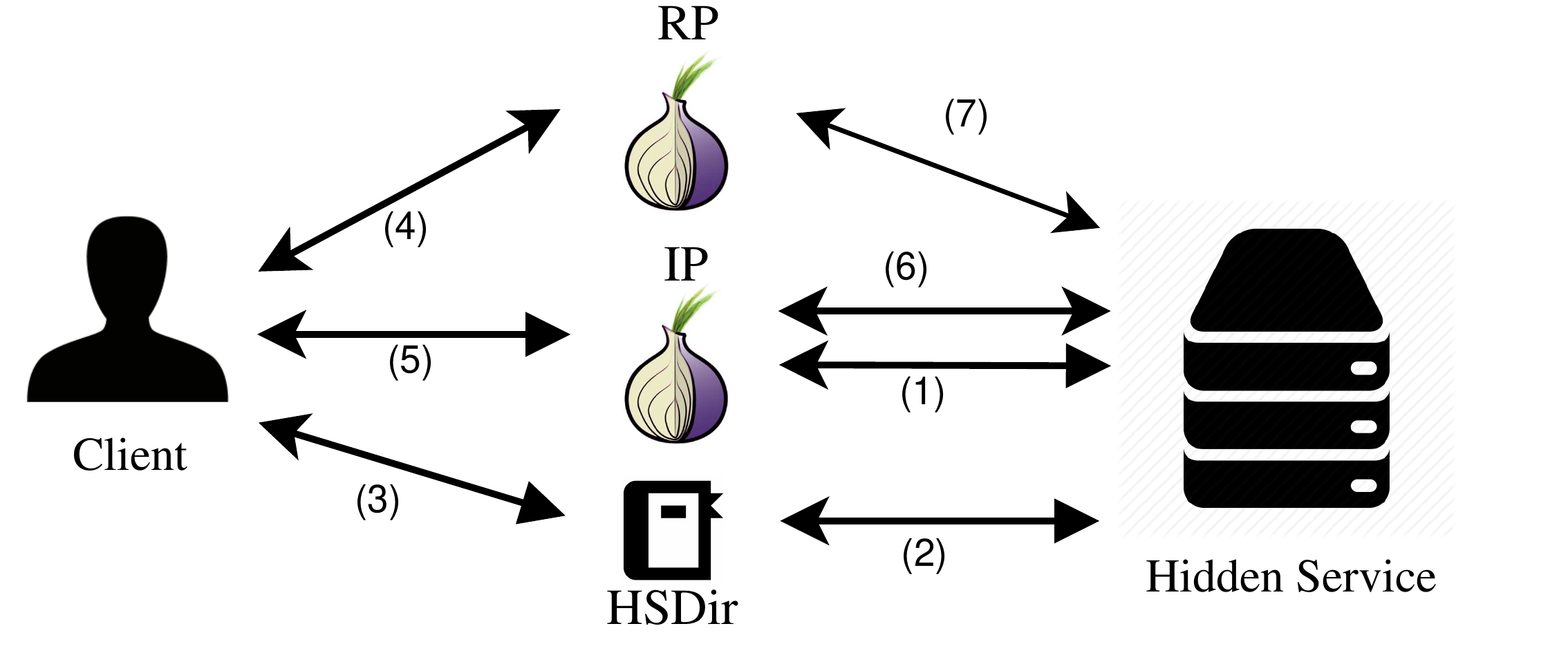}
\caption{Tor hidden service architecture and connection setup.}
\label{f:TorHS}
\end{figure}

\BfPara{Server.} To enable access to a server, the service provider,
generates an RSA key pair. Then he calculates the SHA-1 digest of the
generated public key, known as the \texttt{Identifier} of the hidden
service. The \texttt{.onion} hostname is the base-32 encoding of the
identifier. To connect to a hidden service, the aforementioned
identifier needs to be communicated to the clients through an external
out-of-band channel. As depicted in Figure~\ref{f:TorHS}, the hidden
service, chooses a set of relays, called Introduction Points (IP), and
establishes Tor circuits with them (step 1). After setting up the
circuits, the hidden service calculates two service descriptors to
determine which relays are the responsible HSDirs, using the below formula and uploads the descriptors to them
(step 2).

\ignore{
\noindent
\small \texttt{descriptor\mbox{-}id = H(Identifier~\textbar\textbar~secret-id-part)}\\
\small \texttt{secret\mbox{-}id\mbox{-}part & = & H(time-period~\textbar\textbar~descriptor-cookie \textbar\textbar~replica)}\\
& & \texttt{time\mbox{-}period = (current-time + permanent-id-byte * 86400 / 256) / 86400}
\end{eqnarray*}
}

\texttt{\small\hspace{-1mm}
\begin{eqnarray*}
  \mbox{descriptor-id} & = & \mbox{H(Identifier}||\mbox{secret-id-part)}\\
  \mbox{secret-id-part} & = &
 \mbox{H(time-period}||\mbox{descriptor-cookie} \\ && \mbox{||replica)}\\ 
 \mbox{time-period} & = & (\mbox{current-time} + \\
 && \mbox{permanent-id-byte}
 * 86400 / 256) \\ && / 86400
\end{eqnarray*}
}

\begin{figure}
\centering
\includegraphics[width=0.45\textwidth]{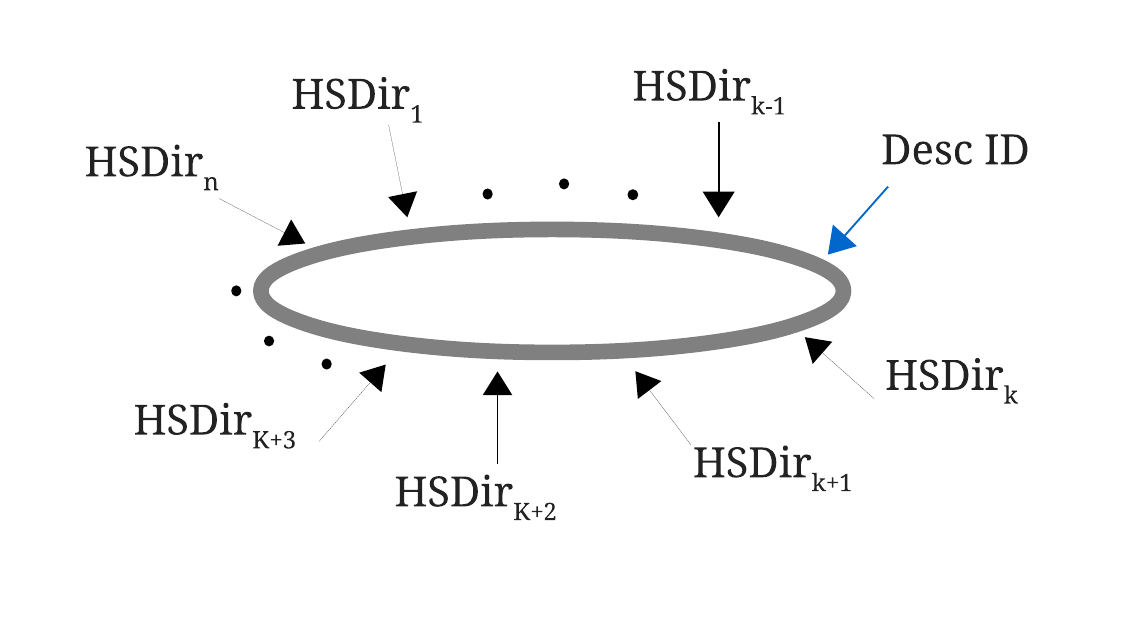}
\caption{Fingerprints circle or Hidden Service Directories (HSDir) and
  placement of a hidden service descriptor. }
\label{f:TorHSDir}
\end{figure}

In the above equations, \texttt{H} is the SHA-1 hash
digest. \texttt{Identifier} is the 80 bit truncated SHA-1 digest of
the public key of the hidden service. \texttt{Descriptor-cookie} is an
optional 128 bit field which could be used for authorization. The
hidden services periodically change their HSDir. The
\texttt{time-period} determines when each descriptor expires and the
hidden services need to calculate the new descriptors and upload them
to the new corresponding HSDirs. To prevent the descriptors from
changing all at the same time, the \texttt{permanent-id-byte} is also
included in the calculations. The \texttt{Replica} index, takes values
of 0 or 1, and results in two descriptors. Each descriptor is uploaded
to 3 consecutive HSDirs, a total of 6. Consider that the circle of
HSDirs is sorted based on their fingerprint (SHA-1 hash of their
public key) as shown in Figure~\ref{f:TorHSDir}. If the descriptor of
a hidden service falls between the fingerprint of HSDir$_{k-1}$ and
HSDir$_{k}$, then it will be stored on HSDir$_{k}$, HSDir$_{k+1}$ and
HSDir$_{k+2}$.

\BfPara{Client.} When a client wishes to contact a hidden service, he
first needs to compute the \texttt{descriptor-id} using the above
formula, and contact the corresponding HSDirs (step 3). To
communicate with a connection with the hidden services, the client
first needs to choose a set of random relays as his Rendezvous Point
(RP), and establish a circuit with them (step 4). Then he contacts the
hidden service's IPs to indicate his desire to contact the hidden
service, and announcing his RPs (step 5). Then, the IP will forward
this information to the hidden services (step 6). At last, the hidden
service establishes a circuit to the RPs, and the two can start
communicating.

\section{Approach}\label{s:approach}

In the following we overview the approach and the architecture of our
detection platform. The steps of flow of actions is depicted in
Figure~\ref{f:honion_architecture}. It consists of the following main
components.

\subsection{HOnions Generation}

In order to automate the process of generating and deploying honions
in a way that they cover a significant fraction of HSDirs, we
developed several scripts. The scripts create configuration files for
Tor relays, called \texttt{torrc}. In particular, the torrc files
specifies the SOCKS port, the hidden service directory to store and
read the private key, the advertised port of hidden service, and the
port where a server is running on the localhost as described in the
next subsection.

A key constraint in this process was to minimize the number of
deployed honions. This derives primarily from our desire to not impact
the Tor statistics about hidden services (specially given the recent
surge anomaly). Secondarily, given that behind each honion there should
be a running process to serve the pages and to log the visits, we are
practically limited by our infrastructure hardware/server
capabilities. We now discuss the process that allowed us to determine
how many honions should be generated to cover at least 95\% of the
HSDir for every batch.

If each honion was only placed on a single random HSDir, the
probability for each HSDir to host an honion is $p_0 =
\frac{1}{N_{hsdirs}}$, where $N_{hsdirs}$ is the number of
HSDirs. Since there are two descriptors, derived independently, this
is equivalent to doubling the number of honions ($m$). Since each
descriptor is placed on a set of three adjacent HSDirs, the
probability of a descriptor being hosted on a HSDir is approximated by
$p \approx 3p_0 = \frac{3}{N_{hsdirs}}$. After generating $m$
honions, the probability that an HSDir is not covered by the $2m$
descriptors is approximated $(1- p)^ {2m}$. To cover a fraction $f$ of HSDir, we need:

\[ f = 1- (1- \frac{3}{N_{hsdirs}})^ {2m} \]

This implies that the necessary honions to be generated should be as follows:

\[ m = \frac{\log(1-f)}{\log(1-(1-\frac{3}{N_{hsdirs}}))}\]

Using this formula and considering that the number of HSDirs
$N_{hsdirs}$ is approximately 3000, we could infer that we need to
generate 1497 (rounded to 1500) honions to cover all HSDirs with
$0.95$ probability. We used 1500 honions per batch (daily, weekly, or
monthly) and could verify that 95\% of the HSDirs were systematically
covered therefore validating our approximation.

An alternative approach would have been to generate a very large
number of honions or interactively generating them until all HSDirs are
covered. However, both approaches have drawbacks and limitations. For
instance, to iteratively cover the HSDirs, one needs to have a
perfect synchronization between the generation process and Tor
consensus documents. As for generating a large number of honions, it
can overload the Tor network, disturb its statistics primitives, and
also requires us to run an excessive number of server processes.

\subsection{HOnion back end servers}

Each honion corresponds to a server process/program that is running
locally. The server behind hidden services, should not be running on a
public IP address. Otherwise it can be detected and deanonymized by
exploiting its unique strings and other
leakages. This has become relatively easy given the
availability of databases of the whole Internet
scans~\cite{durumeric2015search}. To avoid leaking information we
return an empty page for all the services. It does not allow an
adversary to draw any conclusion about the hosting server. We
initially considered using fake pages mimicking real typical hidden
services websites. However, similarities between pages might alert an
adversary about the existence of a honeypot/honey onion.

\subsection{HOnions generation and deployment schedule}

To keep the total number of honions small, we decided on three
schedules for the generation and placement of the honions,
\textit{daily}, \textit{weekly}, and \textit{monthly}. The three
schedules allow us to detect the malicious HSDirs who visit the
honions shortly (less than 24 hours) after hosting them. Since the
HSDirs for hidden services change periodically, more sophisticated
snoopers may wait for a longer duration of time, so they can evade
detection and frame other HSDirs. The daily schedule would miss such
snoopers, therefore we defer to the weekly and monthly honions to spot
such adversaries. Imagine there is a visit on weekly or monthly
honions, while there is no visits to the daily honions. Since all
honions are running simultaneously, and all HSDirs are hosting honions
in all three schedules, this indicates that some malicious HSDirs are
delaying their snooping. For the adversary, this a trade-off between
accuracy and stealthiness, since some hidden services may have a short
life span and will be missed by the snooping HSDir if he waits too
long.

\subsection{Logging HOnions visits}

We log all the requests that are made to the server programs and the
time of each visit. The time of a visit allows us to determine the
HSDirs that have hosted any specific honion. Recording the content of
the requests allows us to investigate the behavior of the snoopers. Since we
advertise our servers on port 80, we can investigate the request types
and content that are made by snoopers. Furthermore, we can detect
automated headless crawls as opposed to the requests made by browsers
(e.g., Tor browser), since they make request for extra elements such
as the small icon that is shown in the browser near the URL address
bar (i.e., favicon.ico).

\subsection{Identifying snooping HSDirs}

Based on the visited hidden server, the time of the visit, and the
HSDir that have been hosting the specific onion address prior to the
visit, we can mark the potential malicious and misbehaving
HSDirs. Then we add the candidates to a bipartite graph, which
consists of edges between HSDirs and the visited honions, as further
described in section~\ref{s:detection}. The analysis of this graph
allows us to infer a lower bound on the number of malicious HSDirs as
well as the most likely snoopers.

\begin{figure*}
\hspace{0.8cm}
\includegraphics[width=0.95\textwidth]{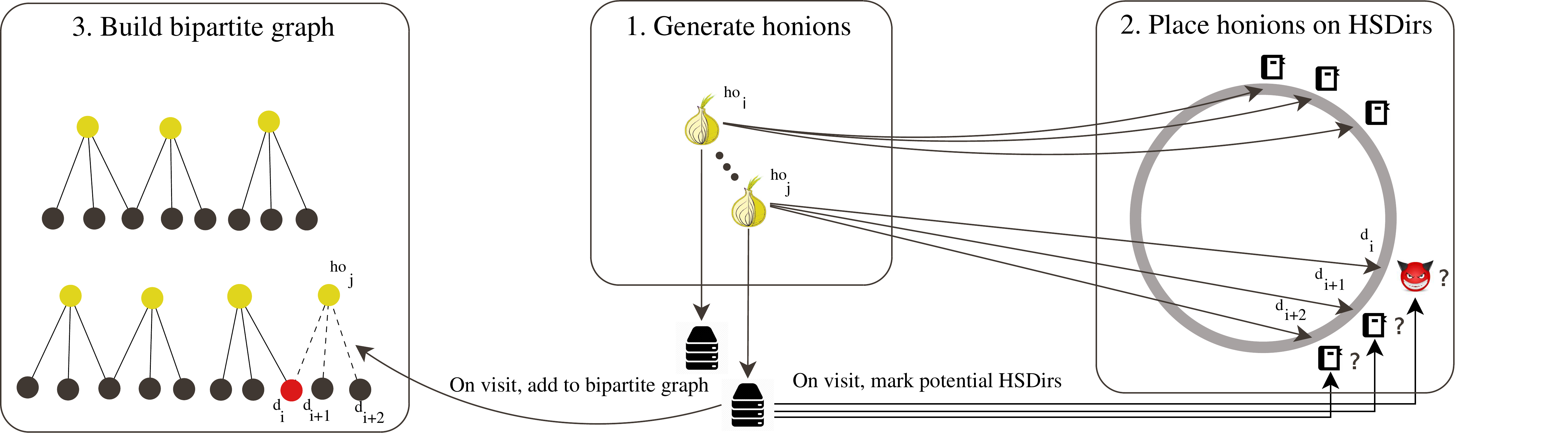}
\caption{Flow diagram of the honion system. We generate a set of
  honions to cover all the HSDirs and run a server behind each one,
  Here, we only show one descriptor per honion. When a visit happens
  to one of the honions, we can infer which HSDirs hosted it (and knew
  about its existence) using the consensus document and the list of
  relays. After identifying the potential suspicious HSDirs, we add
  the candidates to the bipartite graph.}
\label{f:honion_architecture}
\end{figure*}

\section{Estimation \& Identification of Snooping HSDirs}
\label{s:detection}

In order to formally reason about the problem of identifying malicious
HSDirs, we first introduce a formal model and notation for the Honey
Onions system. First, $HO$ denotes the set of honey onions generated
by the system that were visited, and $HSD$ the set of Tor relays
with the \texttt{HSDir} flag (so far referred to as HSDir relays). The
visits of honions allow us to build a graph $G = (V, E)$ whose
vertices are the union of $HO$ and $HSD$ and edges connect a honion
$ho_j$ and HSDir $d_i$ $iff$ $ho_j$ was placed on $d_i$ and
subsequently experienced a visit. $G$ is by construction a bipartite
graph.

\begin{eqnarray*}
HSD &=&  \{ d_i : \mbox{Tor relays with \texttt{HSDir} flag}\} \\
HO &=&  \{ ho_j : \mbox{Honey Onion that was visited}\} \\
V &=&  HSD \cup HO \\
E &=& \{ (ho_j, d_i) \in HO \times HSD | ho_j \mbox{ was placed on } d_i \\
& & \mbox{  and subsequently visited} \} 
\end{eqnarray*}

We also note that each honion periodically changes descriptors and
therefore HSDirs (approximately once a day). However, a HSDir
currently a honion $ho$ cannot explain visits during past
days. Therefore, each time a honion changes HSDirs we clone its vertex
$ho$ to $ho'$ and only add edges between $ho'$ and the HSDirs who know
about its existence when the visit happened.

\subsection{Estimating the number of snooping HSDirs}

Since each honion is simultaneously placed on multiple HSDirs, the
problem of identifying which ones are malicious is not trivial. We
first formulate the problem of deriving a lower-bound on their number
by finding the smallest subset $S$ of $HSD$ that can explain all the
visits (meaning that for each visited honion, there is a member of $S$
who knew about its existence and could therefore explain the
visit). The $S$ is therefore a solution to the following problem:

\begin{equation}
\argmin_{S\subseteq HSD} | S : \forall (ho_j, d_i) \in E\exists
d'_i\in S \wedge (ho_j, d'_i) \in E |
\label{eq:minimal-set}
\end{equation}

The size $s$ of the minimal set tells us that there cannot be less
than $s$ malicious HSDirs who would explain the visits. Furthermore,
when $s$ is relatively small compared to $N_{hsdirs}$, any HSDir
identified as an explanation of multiple visits is highly likely to be
malicious. This derives from the fact that the probability of
co-hosting a honion with a malicious HSDir once being small, it
decreases exponentially as a function of number of visits.

\subsection{Reduction from set cover}

Finding the smallest set $S$ as defined by
Equation~\ref{eq:minimal-set}, is not trivial as one can easily see
that it is equivalent to the hitting set problem, which itself is
equivalent to the set cover problem. The set cover problem is well known
to be NP-Complete. An intuitive sketch of proof for the equivalence to
set cover is as follows. For each HSDir $d_j$ define the set of
honions $O_j = \{ ho_i | (ho_i, d_j) \in E \}$. Solving
Equation~\ref{eq:minimal-set} amounts to finding the smallest set of
$O_j$ that covers all the visited honions. The set cover problem has
an $\ln(n)+1$ approximation algorithm where $n$ is the size of the set
to be covered~\cite{CSLR2001}. Based on this, we derive the following
heuristic, with $\ln(|HO|)+1$ approximation ratio. The advantages of
this heuristic is its low computation complexity $O(|E|)$.

\begin{algorithm}
  \SetAlgoLined
  \KwIn{$G(V, E)$: Bipartite graph of honions to HSDirs}
  \KwOut{S: Set explaining visits}

  $S\leftarrow \emptyset$ \\
  \While{$V\cap HO \neq \emptyset$}{
    Pick $d\in V\cap HSD :$ with highest degree\\
    $V\leftarrow V\setminus\{d \mbox{ and its honion neighbors}\}$\\ 
    }
\caption{Minimal HSDir Heuristic}
\end{algorithm}

\subsection{Formulation as an Integer Linear Program}

Solving the problem defined by Equation~\ref{eq:minimal-set}, can also
be formulated as an Integer Linear Program. Let $x_{1\leq j\leq|HSD|}$
be binary variables taking values 0 or 1. Solving
Equation~\ref{eq:minimal-set}, consists of finding Integer assignments
to the $x_j$ such that:
\[
\begin{array}{ll}
\min_{(x_1, \ldots, x_{HSD})} & \sum_{j=1}^{|HSD|} x_j\\
\mbox{subject to } \forall ho_i\in HO & \sum_{\forall j : (ho_i, d_j) \in E}x_j \geq 1
\end{array}
\]

While this ILP will give the optimal solution, it has exponential
computation complexity in the worst case. In a subsequent section, our
experimental results show that although it performs fairly well for
our setup, it is significantly slower than the heuristic.

\ignore{
\begin{eqnarray*}
\min_{(x_1, \ldots, x_{HSD})} \sum_{j=1}^{|HSD|} x_j\\
\forall ho_i\in HO \sum_{\forall j : (ho_i, d_j) \in E}x_j \geq 1
\end{eqnarray*}
}

\section{Detection Infrastructure \& Results}\label{s:experiment}

In this section we discuss the implementation and deployment of the detection infrastructure as highlighted in Section~\ref{s:approach} and depicted in Figure~\ref{f:honion_architecture}.

\subsection{Implementation and Deployment of the Detection Platform}\label{ss:detect_impl}

We developed simple HTTP servers to listen on specific ports for
incoming requests. Upon receiving a request, each server would log the
time and full request into separate files. At first we developed the
HTTP servers using Python and Flask web framework. However, because of
the size that is occupied by the framework and the interpreter we
faced difficulties in scaling our detection platform. The programs
when instantiated in memory would take up to 40 MB, including the
shared libraries. Running 1500 instances would take up to
12GB. Meaning each instance on average could take about 8-9 MB. As a
result, we decided to port the code to C, without using any external
third party library or framework. We relied solely on the BSD Sockets
API. This allowed us to reduce the size of the code including the
shared libraries to 6 MB. Running 1500 instances with the ported code
only occupied around 2GB, meaning each instance on average occupied
less than 1.5 MB, therefore, reducing the resource allocations by 6
times.

We distributed the 1500 honions over 30 Tor relays equally, to avoid
overloading a single relay and reducing performance and responsiveness
of the hidden services. We created scripts that would automatically
generate and place new honions based on the three schedules discussed
earlier (daily, weekly, monthly). Each schedule was running on a
separate Virtual Machine to isolate the infrastructures.

\subsection{Analyses of the Results and Observations}\label{ss:analyses}

\begin{table}
\centering
\scalebox{1.2}{
\begin{tabular}{ c c c c c }
 \hline
  \textbf{Cloud} & \textbf{Exit} & \textbf{Cloud \& Exit} & \textbf{Not Cloud \& Not Exit}\\ \hline \\
  81  &  27     &  23    & 25  \\ \hline
\end{tabular}
}
\caption{Type of the snooping HSDirs. More than 70\% are hosted on Cloud.} 
\label{t:hsdir-type}
\end{table}

\begin{table}
\centering
\scalebox{0.9}{
\begin{tabular}{ c c c c c }
 \hline
  \textbf{Alibaba} & \textbf{Digital Ocean} & \textbf{Online S.A.S.} & \textbf{OVH SAS} & \textbf{Hetzner Online GmbH}\\ \hline \\
  			15     &  				7     	&  				7   	 & 			6 		&   				6		   \\ \hline
\end{tabular}
}
\caption{Top 5 Cloud Providers.} 
\label{t:cloud-type}
\end{table}

We started the daily honions on Feb 12, 2016; the weekly and monthly
experiments on February 21, 2016, which lasted until April 24,
2016. During this period there were three spikes in the number of
hidden services, with one spike more than tripling the average number of
hidden services (Figure~\ref{f:hs_stat}). First spike was on February
17, second on March 1 (the largest), and the last on March 10. These
spikes attracted a lot of attention from the
media~\cite{bcc-spike}. However, there is still no concrete
explanation for this sharp influx of hidden services. There are some
theories suggesting that this was because of botnets, ransomware, or the
success of the anonymous chat service, called
Ricochet. However, none of these explanations can
definitely justify the current number of hidden services.

\begin{figure}
     \centering
     \subfloat[][Daily Visits]{\includegraphics[scale=0.21]{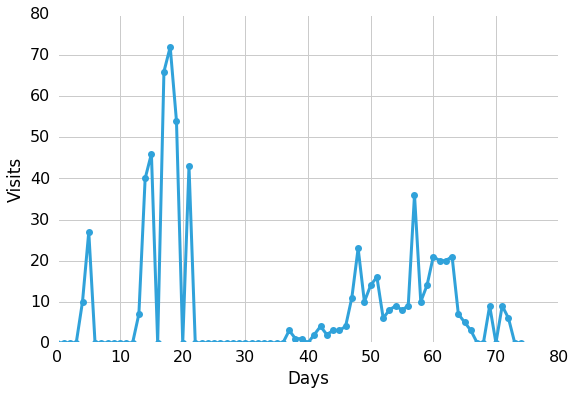}\label{f:daily_visits}}
     \subfloat[][Weekly Visits]{\includegraphics[scale=0.21]{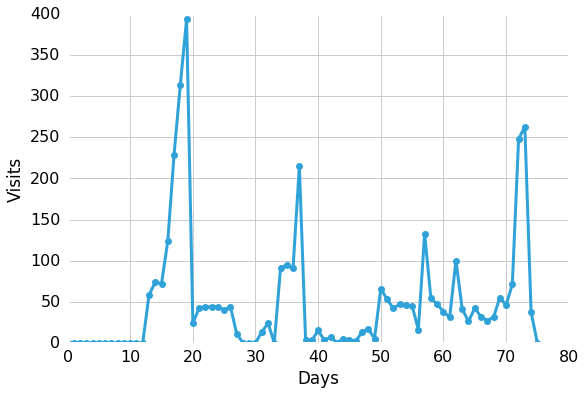}\label{f:weekly_visits}}\\
     \subfloat[][Monthly Visits]{\includegraphics[scale=0.21]{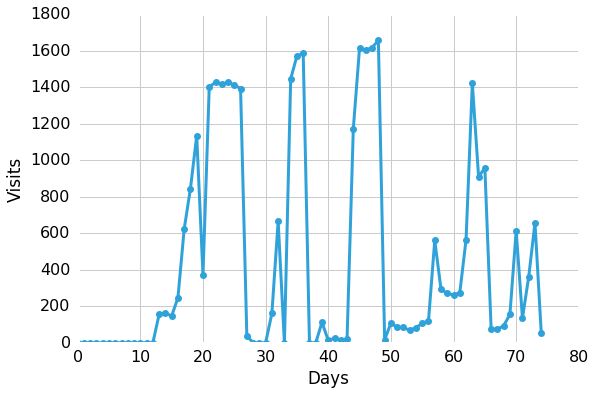}\label{f:monthly_visits}}
     \subfloat[][All Visits]{\includegraphics[scale=0.21]{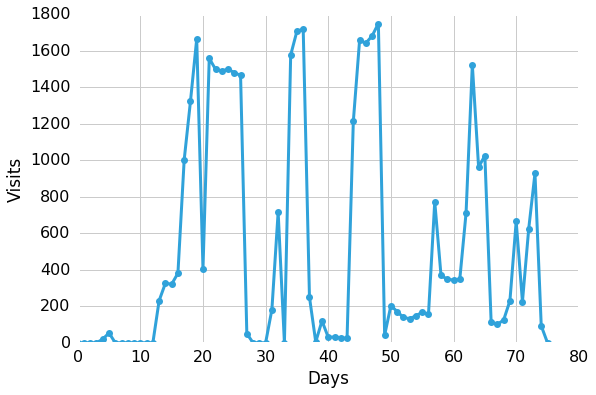}\label{f:all_visits}}
     \caption{Plot of the visits to the honions. The daily onions show
       snooping HSDirs, before the ``mystery'' spike in hidden
       addresses. The number and intensity of the visits is increased
       after the spikes.}
     \label{f:visits}
\end{figure}

Our daily honions spotted snooping behavior before the spike in the
hidden services, this gives us a level of confidence that the
snoopings are not only a result of the anomaly
(Figure~\ref{f:daily_visits}). Rather, there are entities that
actively investigate hidden services. Note that, we started the weekly
and monthly honions after the first spike, even more sophisticated
entities may have been active before, which we are not able to
detect. As we can see in Figure~\ref{f:visits} the visits from
snooping HSDirs increases after the ``mystery'' spikes. Note that, the
delay between the appearance and activity of the snooping HSDirs and
the surge of hidden services, is because of the time it takes for a
relay to acquire the HSDir flag (96 hours and the Stable
flag). Also, whenever a relay gets
restarted it loses its HSDir flags, and they would not see the new
honions, therefore, it introduce further gaps in the daily visits
(Figure~\ref{f:daily_visits}), while the weekly
(Figure~\ref{f:weekly_visits}) and monthly
(Figure~\ref{f:monthly_visits}) visits would still spot activity even
if the HSDir loses its flag.

\BfPara{Snooping HSDirs Nature.} In total we detected at least 110
malicious HSDir using the ILP algorithm (the ILP took about 2 hours), and about 40000 visits. More
than 70\% of these HSDirs are hosted on Cloud infrastructure. Around
25\% are exit nodes as compared to the average, 15\% of all relays in
2016, that have both the HSDir and the Exit flags. Furthermore, 20\%
of the misbehaving HSDirs are, both exit nodes and are hosted on Cloud
systems. The top 5 cloud providers are Alibaba-California (15 detected
HSDirs), Digital Ocean (7), Online S.A.S. (7), OVH SAS (6), and
Hetzner Online GmbH (6). Table~\ref{t:cloud-type} summarizes the cloud
providers and the number of malicious HSDirs each one is
hosting. Alibaba cloud belongs to Alibaba Group, the Chinese
e-commerce company, with servers in the US, Europe and Asia. All
instances that we spotted were hosted on the US West Coast data
centers. Digital Ocean is an American cloud provider that targets
software developers, located in New York. Online S.A.S and Hetzner
Online GmbH are two German cloud provider, and OVH SAS is another
European cloud provider, located in France. Exit nodes play a
significant and sensitive role in the Tor platform, and can cause
legal problems for their operators~\cite{exitproblems}. At the same
time it is known that some Exit nodes are not benign and actively
interfere with users' traffic. There is a \texttt{Bad Exit}
flag, to warn users not to use these relays as exit
nodes. None of the exit nodes that we identified have been identified
as Bad Exit nodes. This can be because they do not perform active MITM
attacks, and evade detection. Table~\ref{t:hsdir-type} summarizes the
type of the HSDir relays.

\begin{figure*}
\centering
\includegraphics[width=0.99\textwidth]{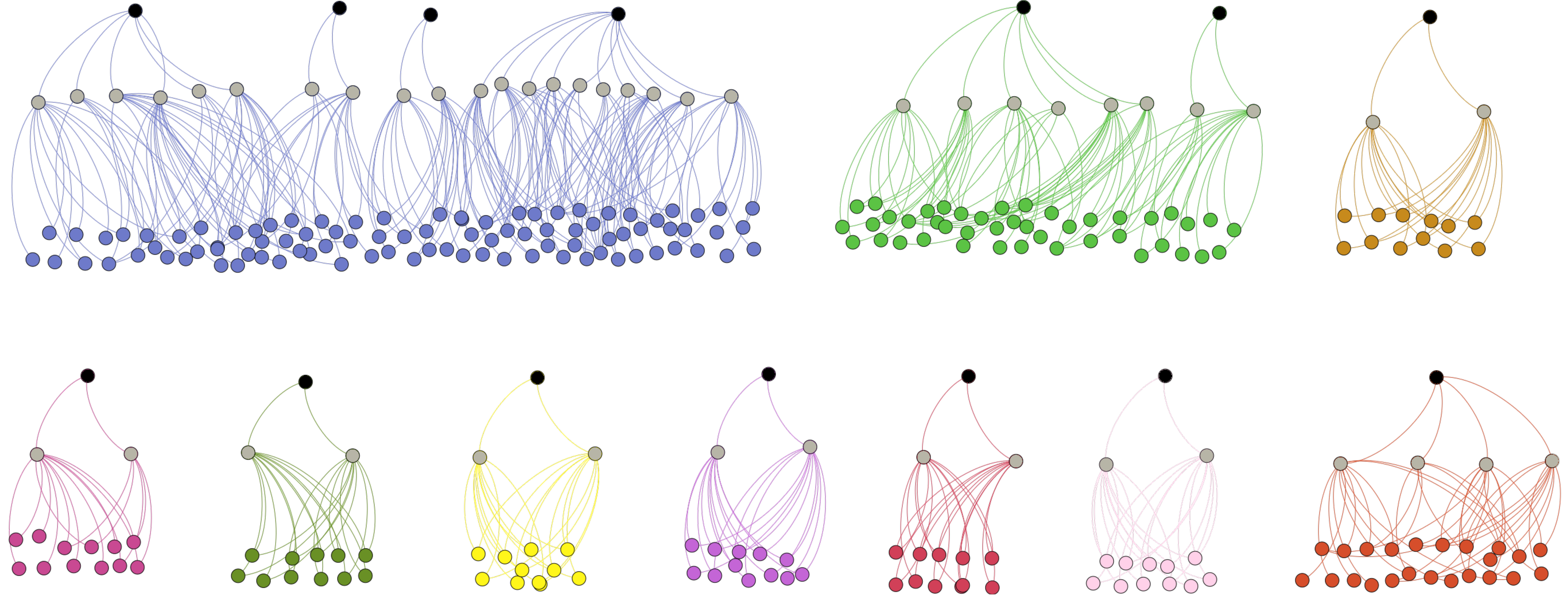}
\caption{A typical graph representation of the visited honions, and the
  hosting HSDirs. The black nodes are the candidate malicious HSDirs
  calculated by the ILP algorithm. The gray nodes are the visited
  honions, and the colored nodes are all the HSDirs that hosted the
  honions, prior to a visit.}
\label{f:onion_hsdir_ilp}
\end{figure*}

Figure~\ref{f:onion_hsdir_ilp} illustrates a typical bipartite graph
of a daily visit. The black nodes indicated the malicious HSDir marked
by ILP. The gray nodes are the honions that have been visited, and the
colored nodes are all other HSDirs that have hosted the honions. Note
that many of the honions belong to a separate component in the graph,
and in one connected component more than one HSDir is suspicious.

\begin{figure}
\centering
\includegraphics[width=0.49\textwidth]{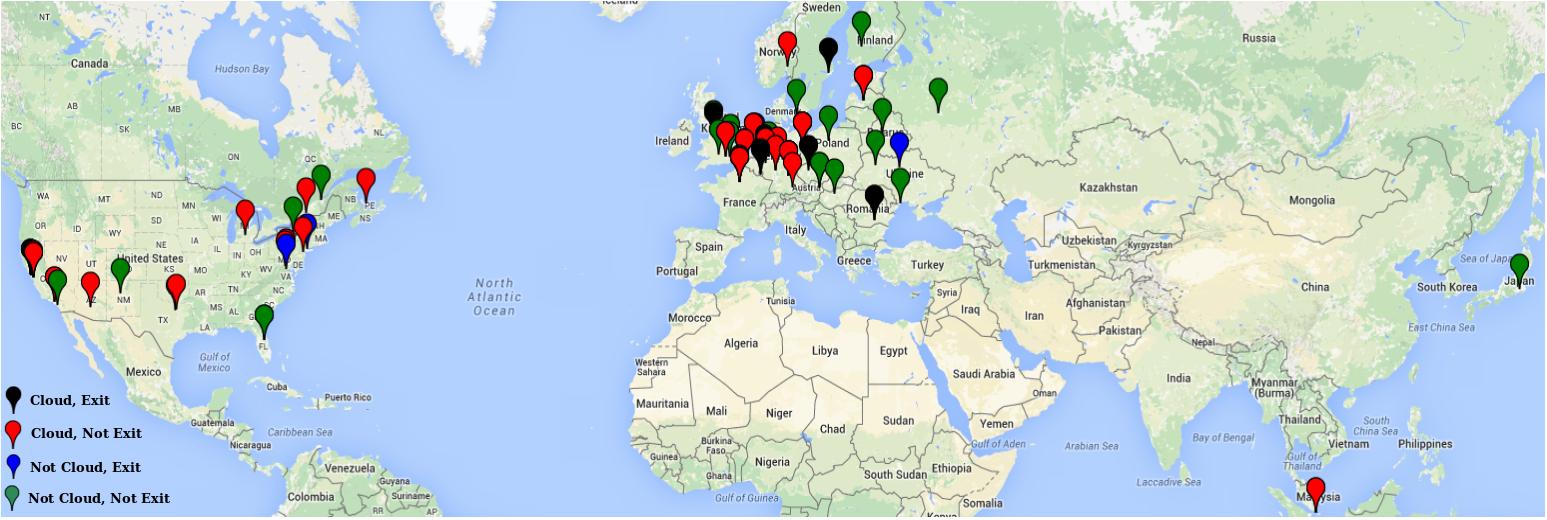}
\caption{The global map of detected misbehaving HSDirs and their
  most likely geographic origin (in the interest of space we have omitted the only
  HSDir in Australia). The black icons represent the HSDirs hosted on
  a cloud platform that are exit nodes as well. The Red icons
  represents the nodes hosted on cloud that are not exit nodes, the
  blue icons represent the exit nodes that are not hosted on the
  cloud, and the green icons are the relays that are neither exit
  nodes, nor hosted on the cloud.}
\label{f:geo_plot}
\end{figure}

\BfPara{Snooping HSDirs Geolocation.} Figure~\ref{f:geo_plot} depicts
the most likely geolocation and type of the misbehaving HSDirs. In the interest of
space we have omitted the only HSDir in Australia. The black icons
represent the HSDirs hosted on a cloud platform that are exit nodes as
well. The Red icons represents the nodes hosted on clouds that are not
exit nodes, the blue icons represent the exit nodes that are not
hosted on the cloud, and the green icons are the relays that are
neither exit nodes, nor hosted on the cloud. Our results indicate that
there are no snooping HSDirs in China, Middle East, or Africa. It is
not surprising since in these regions and countries Tor is heavily
blocked~\cite{censor}. Furthermore, more than 70\% of the snooping
HSDirs are hosted on Cloud systems, and many of the cloud providers'
data centers are located in Europe and Northern
America. Table~\ref{t:hsdir-country} summarizes the top 5 countries
where the malicious HSDirs are located. Note that 15 of the 37 HSDir
in the USA, belong to Alibaba cloud data centers, followed by Digital
Ocean and Linode.

\BfPara{Classifying the Behavior and Intensity of the Visits.} Most of the visits were just querying the root path of the server and
were automated. However, we identified less than 20 possible manual
probing, because of a query for favicon.ico, the little icon that is
shown in the browser, which the Tor browser requests. Some snoopers
kept probing for more information even when we returned an empty
page. For example, we had queries for \texttt{description.json}, which
is a proposal to all HTTP servers inside Tor network to allow hidden
services search engines such as Ahmia, to index
websites. We identified a small number of well behaving civilized
crawler, asking for robots.txt and sitemap. One of the snooping HSDirs
(5.*.*.*:9011) was actively querying the server every 1 hour asking
for a server-status page of Apache. It is part of the functionality
provided by mod\_status in Apache, which provides information on
server activity and performance. This can be an indication of the
adversaries' effort for reconnaissance and finding vulnerabilities and
generally more information about the platform. Tools such as
onionscan~\cite{onionscan} look for such characteristic to ensure
attackers cannot easily exploit and deanonymize hidden services,
because of an oversight in the configuration of the
services. Additionally, we detected different attack vectors, such as
SQL injection, targeting \texttt{the information\_schema.tables},
username enumeration in Drupal (admin/views/ajax/autocomplete/user/),
cross-site scripting (XSS), path traversal (looking for
\texttt{boot.ini} and \texttt{/etc/passwd}), targeting Ruby on Rails
framework (rails/info/properties), and PHP Easter Eggs
(?=PHP*-*-*-*-*).

\begin{figure}
     \centering
     \subfloat[][Contabo GmbH]{\includegraphics[scale=0.30]{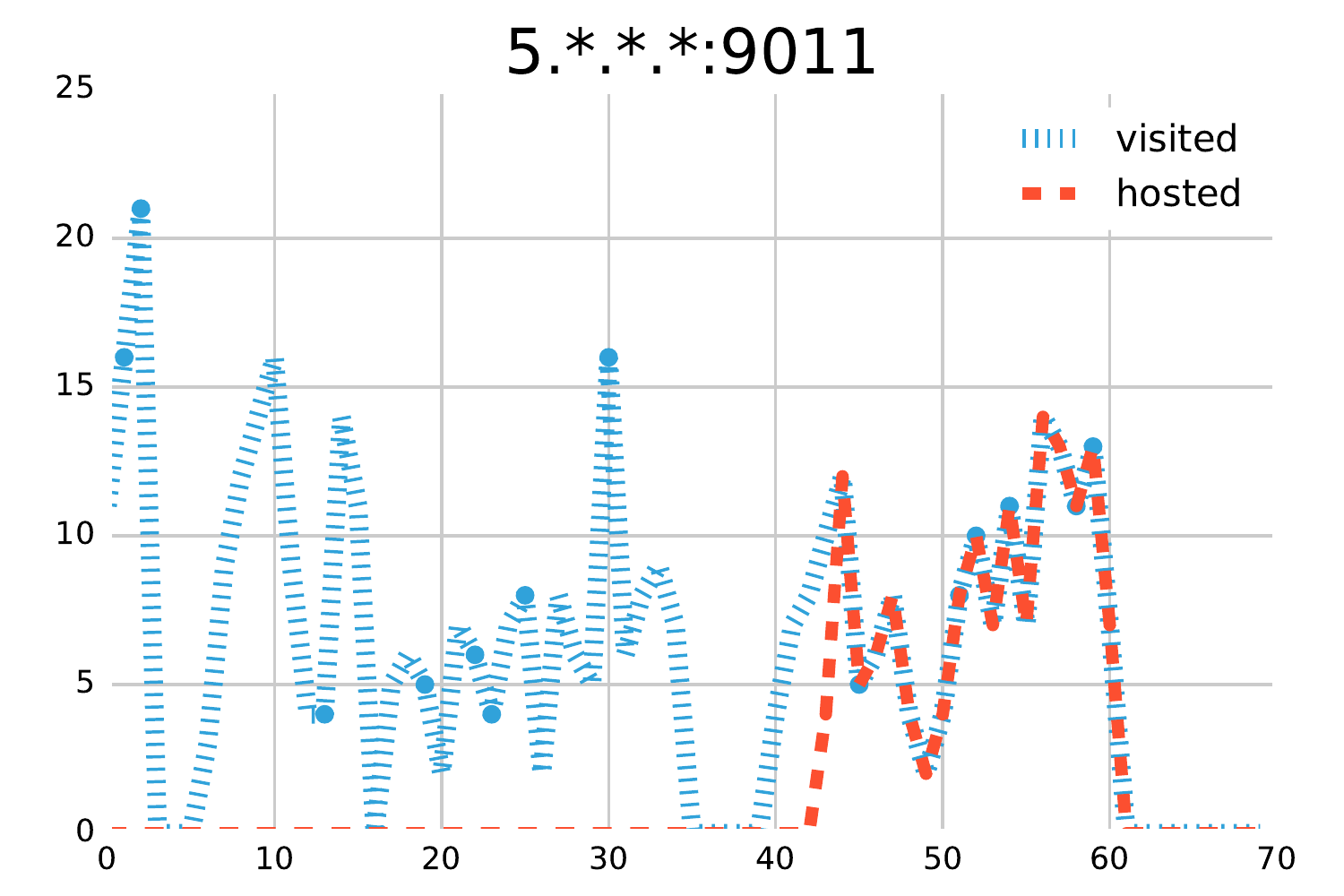}\label{f:german}}
     \subfloat[][Alibaba]{\includegraphics[scale=0.30]{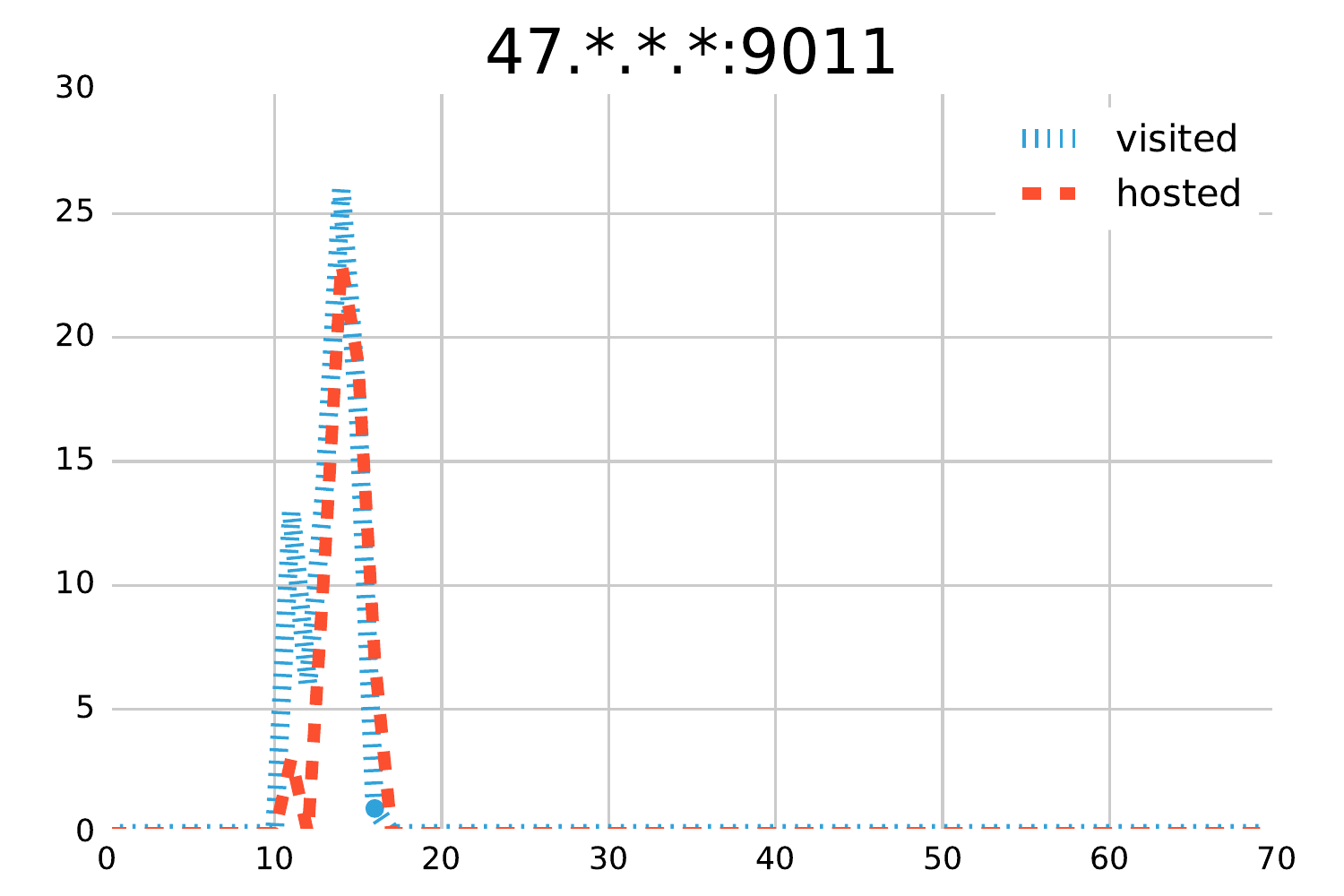}\label{f:alibaba1}}\\
   \subfloat[][Alibaba]{\includegraphics[scale=0.30]{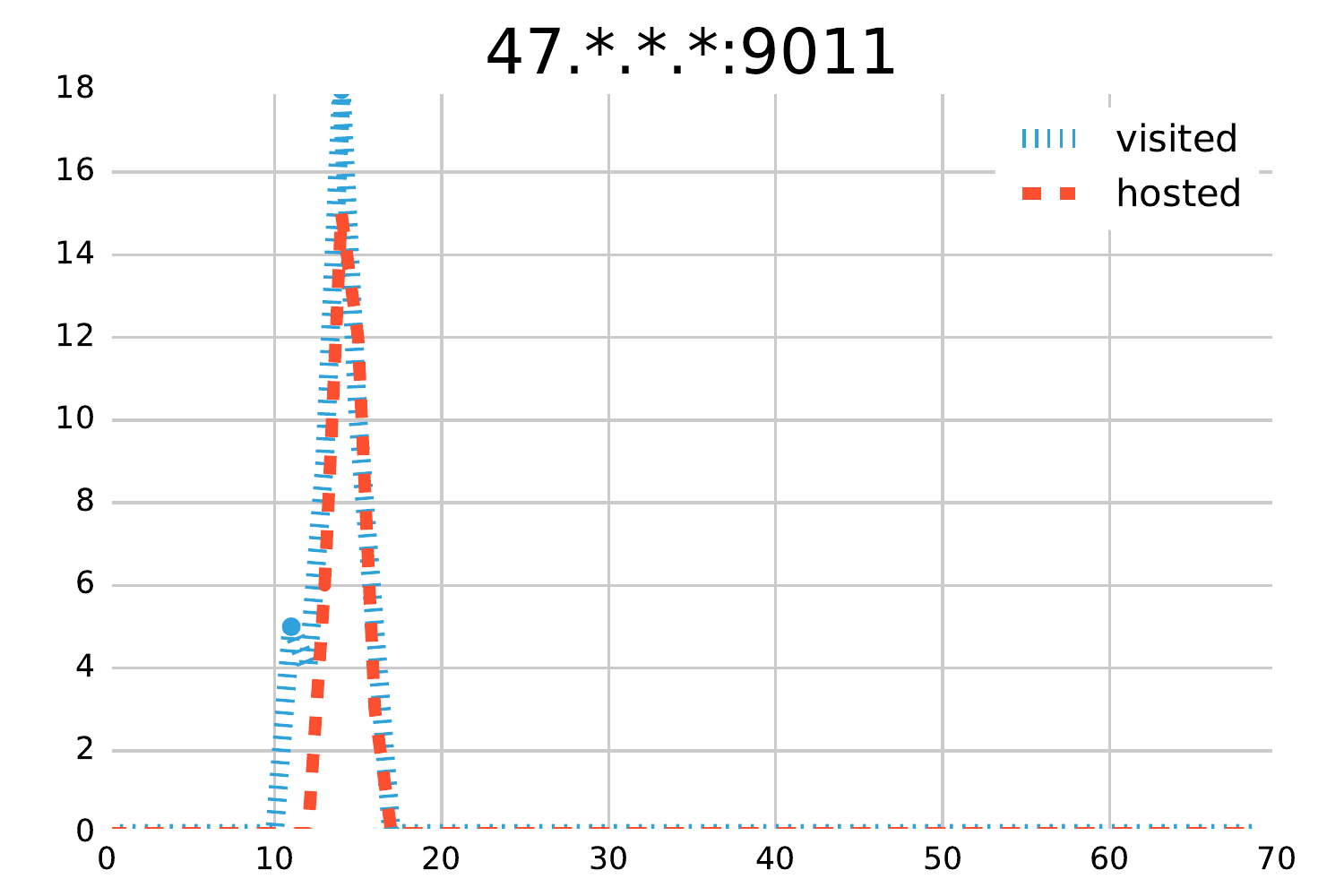}\label{f:alibaba2}}
     \subfloat[][Alibaba]{\includegraphics[scale=0.30]{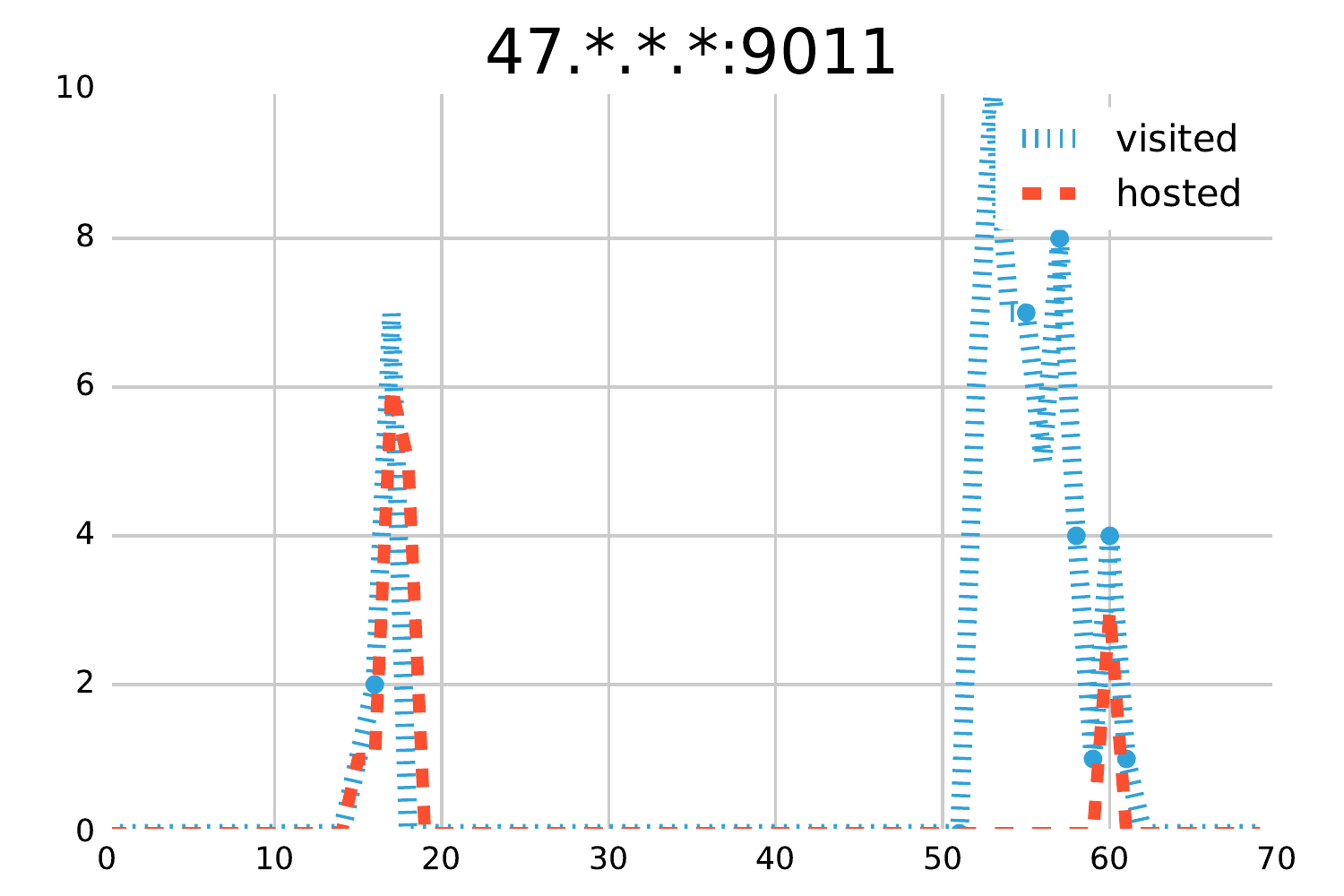}\label{f:alibaba3}}
     \caption{The behavior of the HSDirs. Some of them turned malicious after a while, and some disappeared shortly after the spikes in the number of hidden services.}
     \label{f:hosted_visited}
\end{figure}

In general the snoopers showed a wide range of behavior. Some only
appeared after the first spike in the number of hidden services and
disappeared afterwards and gone offline (Figures~\ref{f:alibaba1}
\&~\ref{f:alibaba2}) (gone offline), while some of them came back
after a month (Figure~\ref{f:alibaba3}). On the other hand, one
snoopers changed its behavior and turned into a snooping HSDirs after
a while (Figure~\ref{f:german}).

\begin{table}
\centering
\scalebox{1.2}{
\begin{tabular}{ c c c c c }
 \hline
  \textbf{USA} & \textbf{Germany} & \textbf{France} & \textbf{UK} & \textbf{Netherlands} \\ \hline \\
  37  &  19     &  14    & 8  & 4 \\ \hline
\end{tabular}
}
\caption{List of top 5 countries with the most likely misbehaving HSDirs.} 
\label{t:hsdir-country}
\end{table}

\section{Discussion \& Future Work}\label{s:discussion}

Based on our observations not all snooping HSDirs operate with the
same level of sophistication. While some do not visit the hosted
honions immediately and therefore evade detection though daily honions,
our weekly and monthly honions can detect them. We believe that
behavior of the snoopers can be modeled and categorized into four
groups. Persistent-Immediate snoopers, where they immediately (within
a day) and systematically probe all .onion addresses they
service. Persistent-Delayed, where they systematically probe all
.onion addresses they service but with a fixed delay d. Randomized
with Deterministic Delay, where they probe a learned .onion address
with probability p after d days. Probabilistic Snoopers, where once
they learn about .onion addresses, they probe, after $d$ days,
according to distribution function $p(d)$. Further work is needed to
define more models and develop techniques to detect and identify the
more sophisticated snoopers.

Since some HSDirs, probe deep in the hidden services, by using
vulnerability discovery and automated attack tools, it would be
interesting to create pages with login forms or more enticing content
to engage the snoopers. However, one should carefully consider the
legal and ethical aspects of such investigations and studies.

The rise and popularity of cloud services allows entities to
provision infrastructures without much overhead, which makes it difficult to
detect malicious Tor nodes. In this competitive market many cloud
providers try to distinguish themselves by providing more privacy and
anonymity for their clients. For example, \texttt{flokinet.is},
advertises its services as a platform suitable for freedom of speech,
investigative journalism, and perfect for whistleblowers, with servers
in Romania, Finland, and Iceland. Although, one can not deny the
benefit of such privacy infrastructures, they can also be subverted
and misused for malicious and harmful activities. Furthermore, cloud
providers such as Vultr, even accepts payments in the form of
bitcoins, which prevents the traceback and identification of
misbehaving entities.

\begin{figure}
\centering
\includegraphics[width=0.45\textwidth]{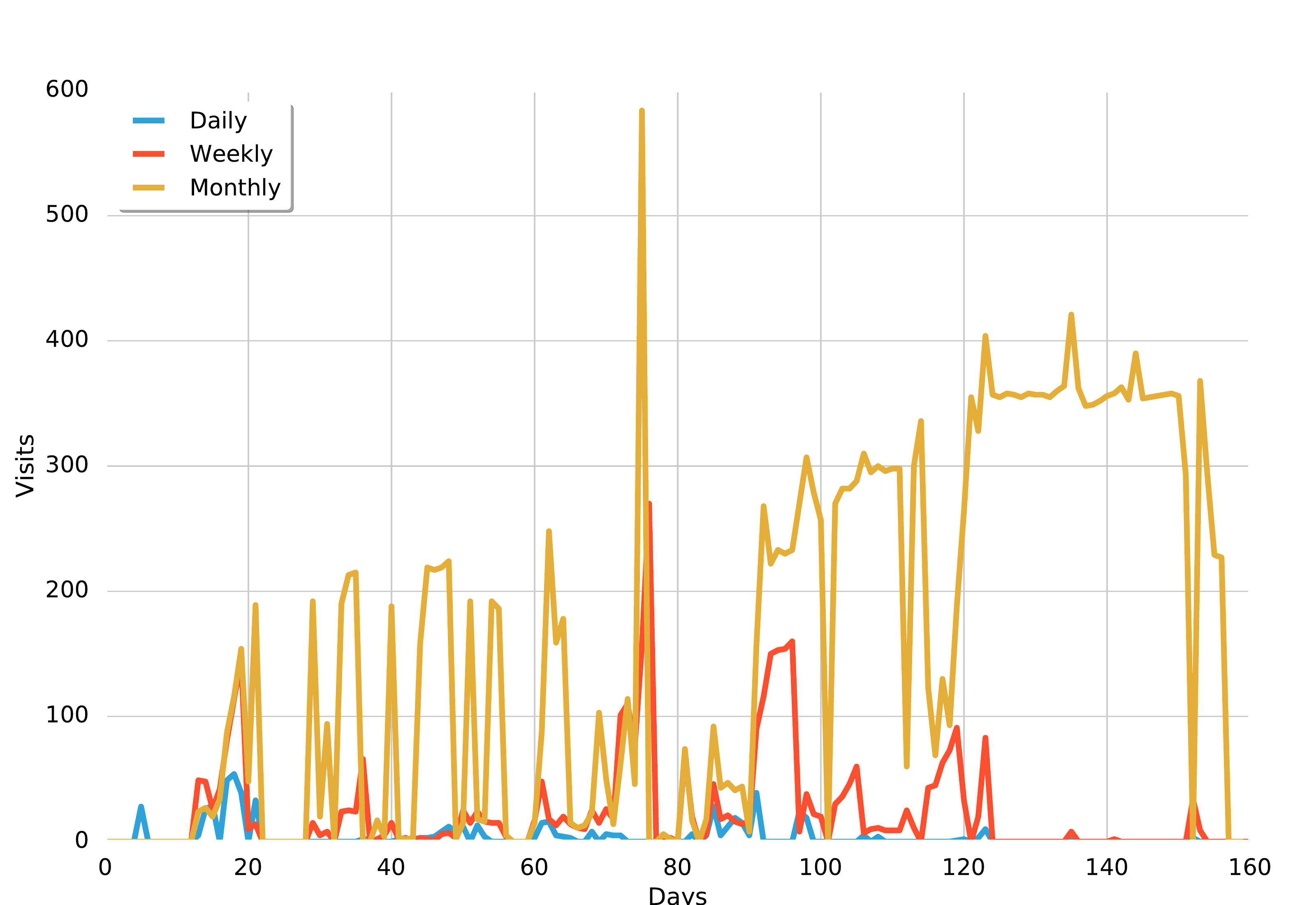}
\caption{The new trend of visits. The snoopers are delaying their visits to avoid identification.}
\label{f:visit_trends}
\end{figure}

It is noteworthy that we continued the deployment of the honions, and after making our work public~\cite{defcon}, we observed a new trend of snooping behavior. The snoopers delay their visits to avoid identification, which indicates that the misbehaving HSDirs have already adapted their techniques. Figure~\ref{f:visit_trends} depicts the new trend of visits, where snoopers are becoming more sophisticated and delay their visits. Note that, we count multiple visits to the same honion within one day, only once for this graph. We also discussed our work with some of the Tor Project people~\cite{donncha} and learned that they have been aware of the problem and developed techniques (although different from ours) to identify  and block misbehaving HSDir relays. Furthermore, they are also  working on a new design to mitigate various attacks against hidden services~\cite{224-rend-spec-ng}. Another direction is to explore the capabilities of Intel SGX~\cite{Jain2016OpenSGXAO,sgx_vb}, and make modifications to Tor to run inside enclaves~\cite{sgxvictor}.

\section{Related Work}\label{s:related}

Previous research studied malicious traffic and misbehaving relays in
the Tor network, however it was mostly limited to the traffic carrying
relays and exit
nodes~\cite{190966,jansen2014sniper,190964,bauer2007low}. Our work
focuses on detection and classification of misbehaving hidden services
directories (HSDirs), an essential component of the hidden services
architecture and the privacy of users. Winter et
al.~\cite{spoiledonion} expose malicious exit nodes by developing two
exit relay scanners, one for credential sniffing and one for active
man-in-the-middle (MITM) attacks. The authors discovered 65 malicious
or mis-configured exit relays participating in different attacks. They
proposed an extension to the Tor browser to thwart MITM attacks by
such malicious exit nodes. In another work~\cite{WinterELF16}, the
authors propose \emph{sybilhunter}, a technique to detect Sybil relays
based on their characteristics such as configuration, fingerprint, and
uptime sequence using the consensus document. Ling et
al.~\cite{torward} present TorWard, a systems for the discovery and
the systematic study of malicious traffic over Tor. The system allows
investigations to be carried out in sensitive environment such as a
university campus, and allows to avoid legal and administrative
complaints. The authors investigate the performance and effectiveness
of TorWard by performing experiments and showing that approximately
10\% of Tor traffic can trigger IDS alert.

Other work explored attacks against Tor to deanonymize users and
hidden services. For example, Biryukov et
al.~\cite{biryukov2012torscan}, document their findings on probing the
network topology and connectivity of Tor relays. The authors
demonstrate how the leakage of the Tor network topology can be
used in attacks to traceback from an exit node to a small set of
possible entry nodes. Therefore, defeating the anonymity of the users
in Tor. In another work~\cite{trawltor}, the authors discover and
exploit a flaw in the design and implementation of hidden services in
Tor, which allows an adversary to measure the popularity of any hidden
service, block access to hidden services, and ultimately deanonymize
hidden services.

Other research looked at the content and popularity of hidden services
and the leakage of .onion address. Biryukovhs et al.~\cite{Biryukovhs} collected 39824 hidden
services descriptors and scanned them for open ports. The author
findings reveal that the majority of hidden services belong to
botnets, followed by adult content and drug markets. Another
study~\cite{thomas2014measuring}, measures the leakage of onion
addresses at the root DNS servers (A and J), and provides the
popularity of different hidden services categories based on the leaked
requests.
\section{Conclusion}\label{s:conclusion}

Tor is a widely popular system for protecting users
anonymity. However, at its core it relies on the non-malicious
behavior of its peer-to-peer nodes. In this work, we introduced honey
onions~\cite{sanatinia-hotpets}, a framework for methodically estimating and identifying the most likely Tor
HSDir nodes that are snooping on hidden services they are hosting. We
propose algorithms to both estimate the number of snooping HSDirs and
identify them. Our experimental results indicate that during the
period of the study (72 days) at least 110 such nodes were snooping
information about hidden services they host. We reveal that more than
half of them were hosted on cloud infrastructure and delayed the use
of the learned information to prevent easy traceback. Another
interesting finding is that although a large number of snooping
HSDirs were hosted on US IP addresses (37), several (15) were actually
hosted on Alibaba's data center in California.
\section{Acknowledgements}\label{s:ack}

This work was supported in part by NSF under grants CNS-1643249.

\bibliographystyle{IEEEtran}
\bibliography{IEEEabrv,bibliography}

\end{document}